\documentclass[%
 reprint,
 amsmath,amssymb,
 aps,
]{revtex4-1}

\usepackage{marginnote}
\usepackage{graphicx}
\usepackage{dcolumn}
\usepackage{bm}
\usepackage{multirow}
\usepackage{rotating}
\usepackage{amsmath,amsfonts}

\usepackage[colorlinks]{hyperref}

\begin{document}

\title{Smooth models for the Coulomb potential}



\author{Cristina E. Gonz\'alez-Espinoza}
\email{gonzalce@mcmaster.ca}
\author{Paul W. Ayers} 
\email{ayers@mcmaster.ca}
\affiliation{%
 Department of Chemistry and Chemical Biology, McMaster University, Hamilton, 1280 Main Street West, L8S 4M1, Canada}
\author{Jacek Karwowski}
\email{jka@fizyka.umk.pl}
\affiliation{%
Institute of Physics, Nicolaus Copernicus University, 87-100 Toru\'n, Poland}
\author{Andreas, Savin}
\affiliation{%
Laboratoire de Chimie Th\'eorique, UMR7616, CNRS, and UPMC, Sorbonne
Universit\'es, F-75252 Paris, France.}
\email{andreas.savin@lct.jussieu.fr}


\begin{abstract}

Smooth model potentials with parameters selected to reproduce the spectrum
of one-electron atoms are used to approximate the singular Coulomb
potential.  Even when the potentials do not mimic the Coulomb singularity,
much of the spectrum is reproduced within the chemical accuracy. For the Hydrogen atom, the smooth
approximations to the Coulomb potential are more accurate for higher angular
momentum states.  The transferability of the model potentials from an
attractive interaction (Hydrogen atom) to a repulsive one (Harmonium and
the uniform electron gas) is discussed.
   
\end{abstract}

\maketitle

%
%

\section{Philosophy}
\label{intro}
How feasible is it to find a model for the Coulomb interaction that is easier
to evaluate but still reproduces key properties of the physical interaction? 
A logical starting point is a system with \textit{no interaction}, as in
Kohn-Sham density functional theory (DFT)\cite{KS65}.  The Kohn-Sham (KS)
approximation starts from a non-interacting system, described as the sum of
the individual electrons' contributions to the energy:
\begin{equation}
\hat{H}_s = \sum_i^N \left( -\frac{1}{2} \nabla^2_i +  v_s(r_i) \right).
\end{equation}

In order to ameliorate the effect of omitting the Coulomb repulsion between the electrons, an extra term, the exchange-correlation functional $E_{xc}[\rho]$, is introduced into the energy expression. For a given external potential $v(r)$

\begin{equation}
E_v[\rho] = T_s[\rho] + J[\rho] + E_{xc}[\rho] + \int v(r) \rho(r) dr 
\end{equation}
where $T_s[\rho]$ is the kinetic energy functional of the non-interacting system, $J[\rho]$ is the classical repulsion, and $E_{xc}[\rho]$ is the exchange-correlation functional, which must be approximated \cite{ParrYang}. The Euler-Lagrange equation associated with the stationary condition of $E_v[\rho]$ can be transformed into a self-consistent set of equations:

\begin{equation}
\left( - \frac{1}{2} \nabla^2 + v(r) + \int \frac{\rho(r')}{|r-r'|}dr' + v_{xc}(r) - \epsilon_j \right) \phi_j (r) = 0
\end{equation}
\[ \rho(r) = \sum_{j=1}^{N} |\phi_j(r)|^2  \]
\[  v_{xc}(r) = \frac{\delta E_{xc}[\rho(r)]}{\delta \rho(r)}. \]

In principle, the KS solutions are exact when $E_{xc}$ is exact, and the KS orbitals  yield the exact density of the system with $N$ electrons in the external potential $v(r)$. The accuracy of KS density functional approximations (DFA) depends on the approximation one uses for $E_{xc}$.
The simplest approximation is the local density approximation (LDA)\cite{LDA80,LDAPZ,LDAPW}. In LDA it is assumed that the exchange-correlation functional is local,
\begin{equation}
E_{xc}[\rho(r)] = \int \epsilon_{xc}(\rho(r)) dr,
\end{equation}
where the exchange-correlation energy density $\epsilon_{xc}(\rho(r))$ at $r$ is taken from the uniform electron gas with density $\rho(r)$.\\

To accurately recover the effect of omitting the interaction between the
electrons, one constructs an adiabatic connection that links the KS
\textit{non-interacting} system with the physical \textit{interacting}
system.  Traditionally, this adiabatic connection is written as a function
of the strength of the interaction, using a simple multiplicative factor
$\lambda$ \cite{HaJo74,LanPer75,GunnLund76,ACYang98}:
\begin{equation}
\hat{H}_{\lambda} = \sum_{i=0}^N -\frac{\nabla_i^2}{2} + v_{\lambda}(r_i) + \frac{1}{2} \sum_{j\neq i} \frac{\lambda}{r_{ij}}.
\end{equation} Computational studies of the adiabatic connection have been performed for few-electron atomic systems, and provide significant insight into the structure of the exact exchange-correlation density functional \cite{Seidl2003,MoriCohenYang06,CohenMoriYang07,ColonnaSavin99,PoColLeinStollSavin03,TeaCorHel09,TeaCorHel10}. 
An alternative to the traditional adiabatic connection is to write the Coulomb interaction as the sum of a short-range piece and a long-range piece. The long-range piece of the potential is usually chosen to be smooth (or at least nonsingular), so that it is relatively easy to approximate solutions to the Schr\"odinger equations when only the long-range piece is included. 

So far we have reviewed traditional strategies that add density-functional
corrections to an ``easy'' system to approximate the real system.  Can we use the real system to construct the model?  For example, is it possible to select an interaction potential, different from the Coulomb one, that nonetheless reproduces a certain target property of
the system?  For example, one might wish to select an interaction potential
that preserves the energy spectrum of an atom.  This strategy is not new. 
Valance and Bergeron \cite{ValBer89} show how to construct analytically
solvable pseudopotentials and model potentials, in the framework of
supersymmetric quantum mechanics, that reproduce experimental spectra. 
Starting from a one-electron one-dimensional Hamiltonian $H_1$ associated
with the potential $V_1$, they found a supersymmetric partner $H_2$,
characterized by a second potential $V_2$, with almost the same spectrum
as $H_1$.  $H_2$ is missing the ground-state of $H_1$.  A similar approach
has been used by Lepage\cite{Lep97} in the field of elementary particle
physics, where the Hamiltonian is constructed to reproduce low-energy
features of a particular physical system.

In the next section we define an expression for the model potential. We then
explain two different model potentials that accurately reproduce the
lowest-energy eigenvalues of the Hydrogen atom.  In section 4, we use the
same models for the Coulomb potential to replace the Coulomb repulsion
between the electrons in two-electron Harmonium.  Finally, the exchange energy of the uniform electron gas that results from one of the models is compared to the standard approximation.

\section{Ansatz}
\label{sec:1}
Analogous to the inverse problem of finding the Kohn-Sham potential from a
given density, where oscillatory potentials and/or shifted potentials can
reproduce the exact density numerically \cite{SavinColPol03}, finding the potential given the spectrum is not trivial because the solution is not unique.  Therefore, we restrict the analytical form of the potential by imposing some constraints.  We would like to eliminate the singularity of
the Coulomb potential because solving the Schr\"odinger equation for a singular operator is computationally demanding.  We also wish to preserve the long-range asymptotic form of the potential, so that the long-range
electrostatics is correct.  An interaction potential that satisfies
these constraints is:
\begin{equation}
\frac{1}{r} \rightarrow V_\mu(r)=
c\,\exp(-\alpha^2 r^2)+\frac{\mathrm{erf}(\mu r)}{r}.
\label{eq_pot}
\end{equation}
We prove in this article that despite the simplicity of this
\textit{erfgau} \cite{TouSavinFlad} type of potential, it is flexible enough
for our purposes.

\section{H atom}
\label{sec:2}
\subsection{Construction of $V_\mu(r)$}

To determine the parameters in the model potential we consider what happens
when we replace the Coulomb interaction between the nucleus and the electron
in the hydrogen atom by the model potential $V_\mu(r)$ (\ref{eq_pot}). Thus, 
we replace the Hamiltonian of the Hydrogen atom
\begin{equation}
\hat{H}_0(\mathbf{r}) = -\frac{1}{2}\nabla^2 - \frac{1}{r}
\end{equation} 
by a modified Hamiltonian
\begin{equation}
\hat{H}_\mu(\mathbf{r})=-\frac{1}{2}\nabla^2-V_\mu(r)
\label{hamiltonian}
\end{equation}
with $V_\mu(r)$ defined in such a way that 
\begin{equation}
\label{lim}
\lim_{\mu\to\infty}\hat{H}_\mu=\hat{H}_0,\;\;\;\mbox{i.e.}\;\;\;
\lim_{\mu\to\infty}V_\mu=\frac{1}{r}.
\end{equation}
As we know, in the case of the long-range term
\[
\frac{\mathrm{erf}(\mu\,r)}{r}
\genfrac{}{}{0pt}{2}{\textstyle{\sim}}{\mu\to\infty}\,\frac{1}{r}.
\]
Thus, condition (\ref{lim}) is fulfilled if
$[c\,\exp(-\alpha^2 r^2)]\to 0$ when $\mu \to \infty$. Besides, we would
like the spectrum of $\hat{H}_\mu$ to be as close as possible to the
spectrum of $\hat{H}_0$.  This can be achieved by properly choosing 
$c=c(\mu)$ and $\alpha=\alpha(\mu)$.  

Consider $\hat{H}_0$ as the unperturbed operator, and
\begin{equation}
w_\mu=\hat{H}_\mu-\hat{H}_0
\label{pert}
\end{equation}
as a perturbation. First, we notice that for the bound states of the Hydrogen atom
\begin{equation}
\langle \psi_i|\mathrm{erfc}(\mu\,r)/r|\psi_i\rangle\,
\genfrac{}{}{0pt}{2}{\textstyle{\sim}}{\mu\to\infty}\,\mu^{-(2l+2)},
\label{asymp1}    
\end{equation}
and
\begin{equation}
\langle \psi_i|c\,\exp(-\alpha^2r^2)|\psi_i\rangle\,
\genfrac{}{}{0pt}{2}{\textstyle{\sim}}{\alpha\to\infty}\,
c\,\alpha^{-(2l+3)},
\label{asymp2}  
\end{equation}
where $l$ is the angular momentum quantum number, a necessary condition for the spectra of $\hat{H}_\mu$ and $\hat{H}_0$ to coincide is that these two expectation values have the same asymptotic form. (See Appendix A for more details about the $\mu$-dependence of the wavefunction.) The simplest choice, adopted in this paper, is to take $\alpha$ as a linear function of $\mu$. This implies that $c$ also has to be linear in $\mu$. Thus, we can set
\begin{equation}
c=\gamma\,\mu,\quad \quad \alpha=\kappa\,\mu,
\label{param}
\end{equation}
where $\gamma$ and $\kappa$ are $\mu$-independent parameters. 

\begin{figure}[h]
  \includegraphics[width=0.5\textwidth]{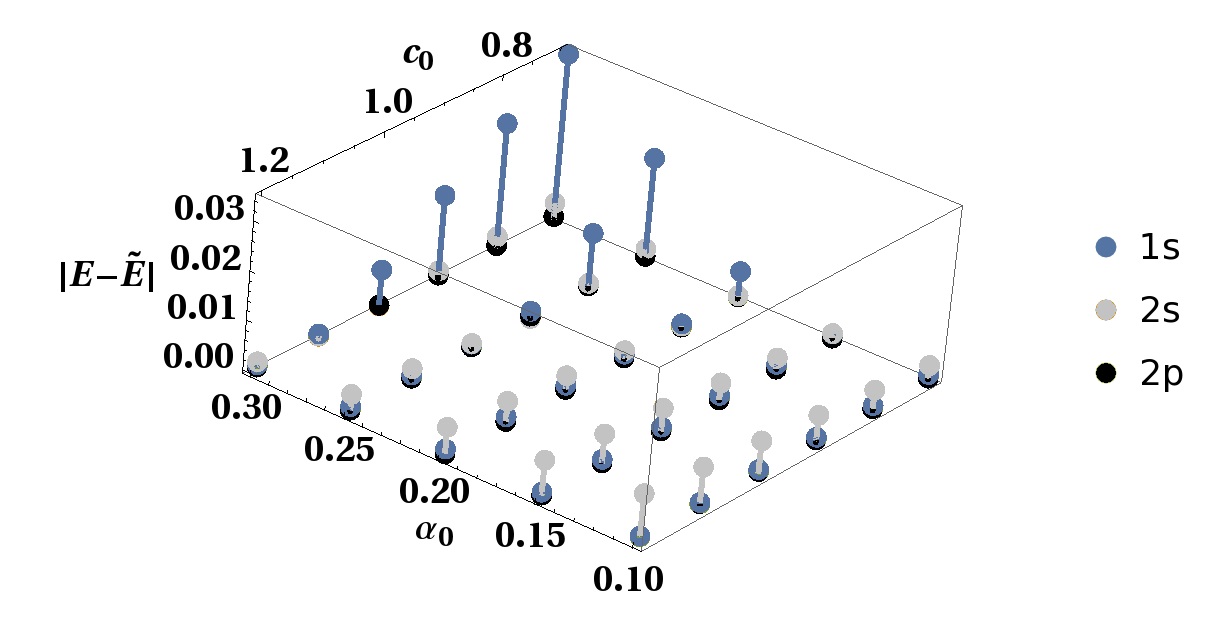}
\caption{The accuracy of the model potential spectrum with respect to the 
parameters $c_0$ and $\alpha_0$. Energy errors (in \%), for $\mu = 1$.}
\label{fig_ca}
\end{figure}

According to Eqs. (\ref{asymp1}), (\ref{asymp2}) and (\ref{param}), the asymptotic expansion of the expectation value of $w_\mu$ may be written as
\begin{equation}
\langle\psi_i|w_\mu|\psi_i\rangle=\sum_{j=2l+2}d_j(\gamma,\kappa)\,
\mu^{-j},\quad \mu>>1.
\label{pt1st}
\end{equation}
We select $\gamma$ and $\kappa$ so that the two leading terms in expansion (\ref{pt1st}) vanish. 
The lowest order terms ($j=2$ and $j=3$) correspond to $l=0$
states.  For angular momenta $l>0$ the leading terms in
Eq.~(\ref{pt1st}) are $O(\mu^{-4})$ or smaller.  Then, the  
choice of $c(\mu)$ and $\alpha(\mu)$ is dictated by the requirement that the eigenvalues of the $S$ states are as correct as possible for $\mu \to \infty$.  The
explicit form of expansion (\ref{pt1st}) for $l=0$ states reads
(see Appendix B)
\begin{equation}
\small
\label{taylor4th}
\langle\psi_i|w_\mu|\psi_i\rangle=
\left(1-\frac{\sqrt{\pi}\gamma}{\kappa^3}\right)\mu^{-2}  
+\left(-\frac{8}{3\sqrt{\pi}}+\frac{4\gamma}{\kappa^4}\right)\mu^{-3}
+O(\mu^{-4}).
\end{equation} 
Solving equations $d_2=d_3=0$ for $\gamma$ and $\kappa$, we obtain:
\begin{equation}
c =\frac{27}{8\sqrt{\pi}}\,\mu \, = 1.904 \, \mu, \; \; \;\; \alpha =\frac{3}{2}\,\mu \,= 1.500 \, \mu.
\label{eq_as}
\end{equation}
In addition to the asymptotic behavior, for practical calculations we need the optimal parameters $c$ and $\alpha$ at finite values of $\mu$. Taking the linear forms
\begin{equation}
\label{lin-form}
c=\gamma\,\mu+c_0,\quad \quad \alpha=\kappa\,\mu+\alpha_0
\end{equation} 
with the parameter $\gamma$ and $\kappa$ from the previous step we can use $c_0$ and $\alpha_0$ to further optimize the spectrum. 
The condition for the elimination of $\mu^{-2}$ term remains the same as before. The equations $d_3=0$ and $d_4=0$ are given 
in Appendix B.  The behavior of the spectrum of the model potential versus
$c_0$ and $\alpha_0$ is shown in  Figure~\ref{fig_ca}, where for fixed
$\mu=1.0$ energies of 1s, 2s and 2p states are displayed. As one 
can see, there is a range for which pairs of $(c_0,\alpha_0)$
give reasonably small errors of the energy values. The dependence of the relative error on $n$ and $l$ is discussed in section~\ref{sec_eigenvalues}.

To select the best linear forms of $c$ and $\alpha$ we constructed a grid on the
intervals $c=[-0.5, 0.0]$ and $\alpha=[1.0, 3.0]$, and then we computed the
error in the eigenvalues of the 1s, 2s, 2p, 3s, 3p, and 3d states for
$\mu=[0.5,2.0]$. We defined the best choice for the parameters as the minimax choice: the $c,\alpha$ that minimized the maximum absolute deviation between the eigenvalues with the model potential and the exact result from the Coulomb interaction,
\begin{equation}
\delta=\min_{c(\mu),\alpha(\mu)}\{\max_{n,l}|
E_\mathrm{Coulomb}-E_\mathrm{model}|\}.
\label{eq_minmax}
\end{equation}
\begin{figure}[h]
 \includegraphics[width=0.5\textwidth]{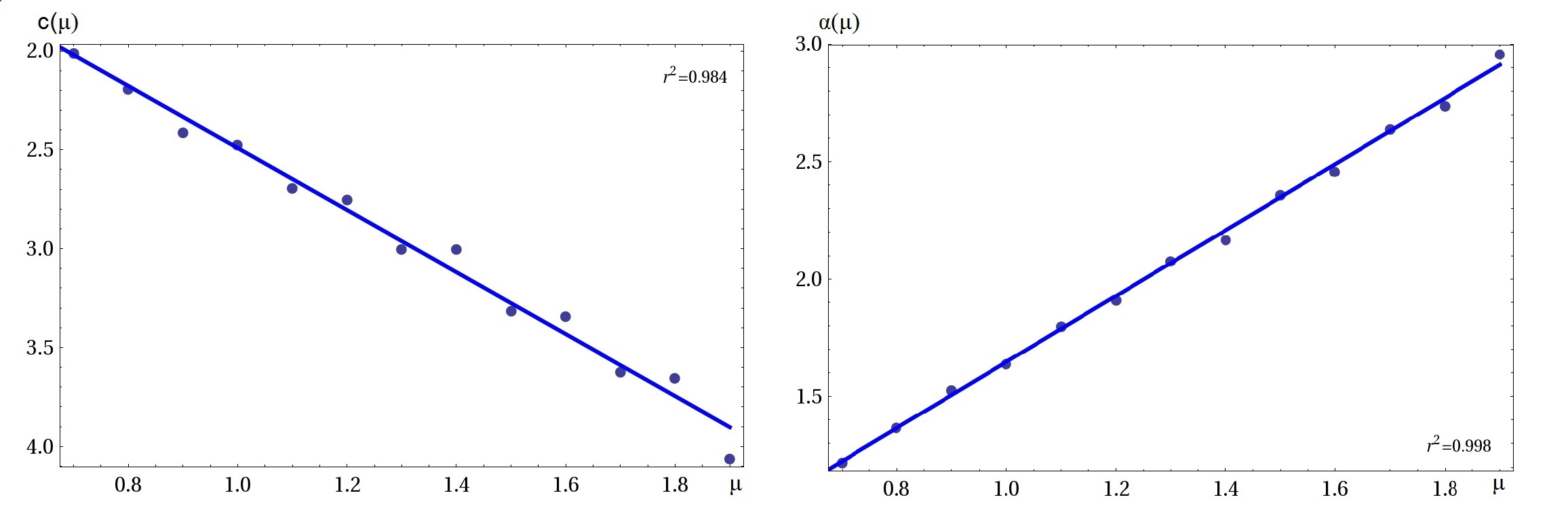}
\caption{Linear regressions for the two parameters $c(\mu)$ and $\alpha(\mu)$ 
of the simple fit in Eq.~(\ref{eq_fit}). Here, the dots are the optimal values of the parameters according to Eq.~(\ref{eq_minmax}), the lines are the least-squares linear regressions, and $r^2$  are their corresponding coefficients of determination.}
\label{fig_fit}
\end{figure}
As seen in Figure~\ref{fig_fit}, the best values of $(c,\alpha)$ can be 
modeled by a linear function,
\begin{align}
c&=0.923+1.568\,\mu\nonumber \\ 
\alpha&=0.241+1.405\,\mu.
\label{eq_fit}
\end{align}
Note that the resulting fit is very similar to one the linear forms obtained from perturbation theory \{$c_0=0.943$, $\gamma = 1.904\, \mu$\} and \{$\alpha_0=0.247$, $\kappa = 1.5\,\mu$\} (see equation~(\ref{eq_pt-lineal}) of Appendix B). With this fitted form the interaction does not vanish when $\mu=0$, therefore the spectrum cannot be exact for small values of $\mu$, in contrast to the asymptotic form (\ref{eq_as}), in which both $c$ and $\alpha$ are proportional to $\mu$.

The optimum parameters for a Hydrogen-like atom with the nuclear charge $Z$
may be obtained from the ones determined for the case of $Z=1$ by a simple 
scaling procedure. Eq. (\ref{eq_pot}) becomes
\begin{equation}
\frac{Z}{r}\rightarrow V^Z_\mu(r)=
Z\left[c_Z\,\exp(-\alpha_Z^2 r^2)+\frac{\mathrm{erf}(\mu_Z r)}{r}\right]
\label{eq_pot_z}
\end{equation}
and
\begin{equation}
\label{hamiltonian-z}
\hat{H}_\mu^Z(\pmb{r})=-\frac{1}{2}\nabla^2-V_\mu^Z(r)=
Z^2\,\hat{H}_\mu(\pmb{\rho}),
\end{equation}
where $\pmb{\rho}=Z\,\pmb{r}$ and
\begin{equation}
\label{scaling}
c_Z=Z\,c,\quad\alpha_Z=Z\,\alpha,\quad\mu_Z=Z\,\mu.
\end{equation}

\subsection{Results}

\subsubsection{Potentials}
\begin{figure}[h]
 \includegraphics[width=0.5\textwidth]{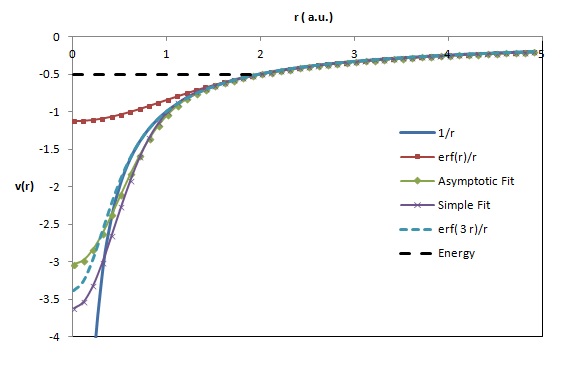}
 \includegraphics[width=0.5\textwidth]{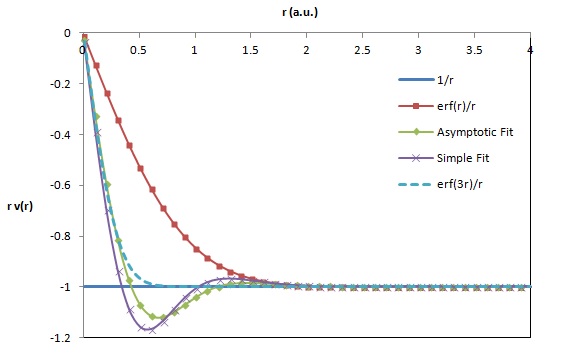}
\caption{Comparison between the Coulomb potential (solid line) and the
model potentials: long-range(squares), asymptotic (diamonds), fitted (stars)
and modified long-range potential (dashed line).}
\label{fig_v}
\end{figure}

The model potentials we consider in this paper [$\mathrm{erf}(\mu r)/r$, 
asymptotic (Eq. (\ref{eq_as})), and fitted (Eq. (\ref{eq_fit})] corresponding to $\mu=1$ are compared with the
Coulomb potential and with a modified long-range potential,
$\mathrm{erf}(3r)/r$, in Figure~\ref{fig_v}.
One might suspect that adding an optimized Gaussian term to
$\mathrm{erf}(\mu r)/r$ would give a potential that mimics the effect of
increasing $\mu$ in the long-range term.  This is not the case for the
potentials we consider in this paper.  At first glance the potential
$\mathrm{erf}(3 r)/r$ (dashed line) seems similar to the asymptotic and to
the fitted potentials.  But the $\mathrm{erf}(3r)/r$ potential is {\em always} weaker than the
Coulomb potential, while the latter potentials, though in some intervals of $r$ they are also weaker, in other intervals they are stronger than the Coulomb potential. 
This may explain why the asymptotic and fitted potentials reproduce the
spectrum much better than the modified long-range potential: the effects of
too strong and too weak regions of the model potentials cancel each other,
leaving the eigenvalues relatively unchanged.

\subsubsection{Eigenvalues} \label{sec_eigenvalues}

\begin{figure}[h]
   \includegraphics[width=0.5\textwidth]{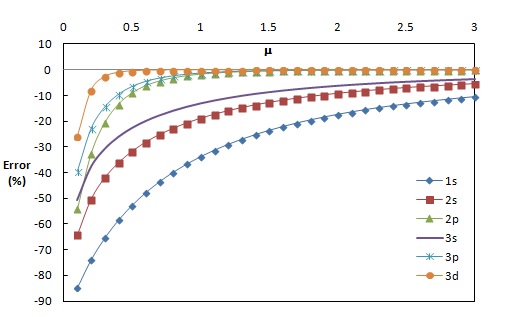}
   \includegraphics[width=0.5\textwidth]{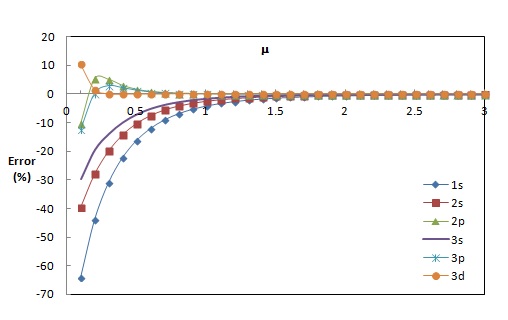}
   \includegraphics[width=0.5\textwidth]{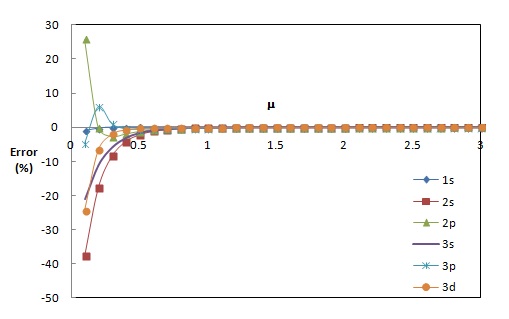}
\caption{Percentage errors in the eigenvalues of Hydrogen. From top to
bottom: a)long-range $\mathrm{erf}(\mu r)/r$, b) asymptotic
[Eq.~(\ref{eq_as})], and c)fitted [Eq.~(\ref{eq_fit})] potentials.}
\label{fig_h-percent}
\end{figure}

\begin{figure}[h]
   \includegraphics[width=0.5\textwidth]{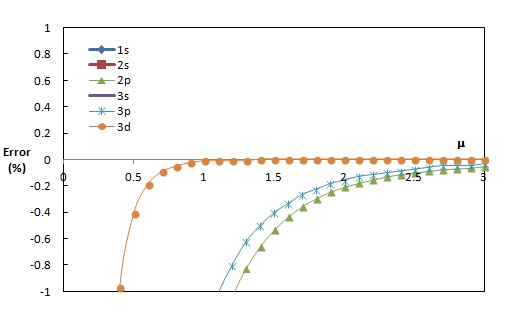}
   \includegraphics[width=0.5\textwidth]{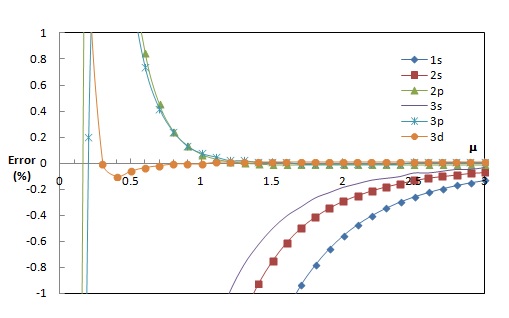}
   \includegraphics[width=0.5\textwidth]{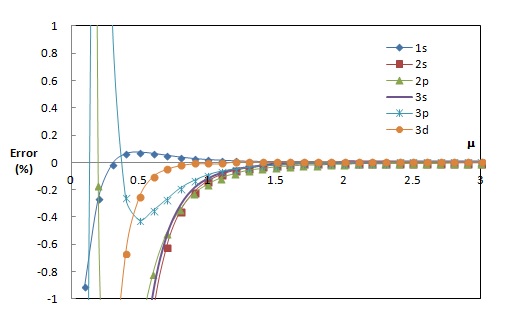}
\caption{Close-up of the percentage error of the eigenvalues of Hydrogen. As
in Figure~\ref{fig_h-percent}, the curves are, from top to bottom:
a)long-range $\mathrm{erf}(\mu r)/r$, b) asymptotic [Eq.~(\ref{eq_as})], and
c)fitted [Eq.~(\ref{eq_fit})] potentials.}
\label{fig_h-blowup}
\end{figure}

The percentage errors in the eigenvalues of Hydrogen with long-range,
asymptotic, and fitted potentials are presented in
Figure~\ref{fig_h-percent}.  As the quantum number increases, the amplitude
of the eigenfunctions near the nucleus decreases, the long-range part of the
potential dominates, and the eigenvalues approach the exact ones. A
clear improvement is found in the asymptotic and fitted potentials compared
with the traditional long-range potential.  As expected, the fitted
potential produced the smallest errors. However, for $\mu > 1.5$, the difference
between the asymptotic and the fitted models is rather small; see
Figure~\ref{fig_h-blowup}.

\begin{figure}[h]
   \includegraphics[width=0.5\textwidth]{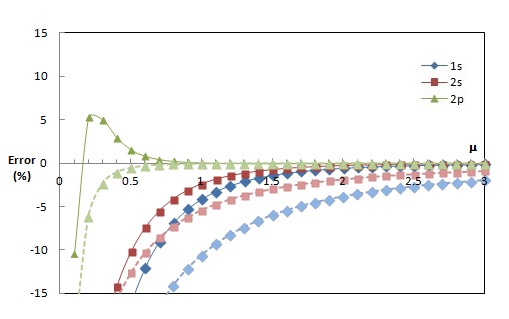}
\caption{Comparison of the percent error in the first three eigenvalues of the Hydrogen
atom between the asympotic potential, (thick lines) and $\mathrm{erf}(3
\mu)/r$ (dashed lines).}
\label{fig_h-erf3}
\end{figure}

The advantage of adding a Gaussian term is clear when we compare against a
modified long-range potential, such as $\mathrm{erf}( 3 \mu r)/r$,
Figure~\ref{fig_h-erf3} .  The Gaussian term lets us ``get away with" a much
smaller value of $\mu$, and seems to work better for s-type orbitals than the erf-based potential.

\begin{figure}
   \includegraphics[width=0.5\textwidth]{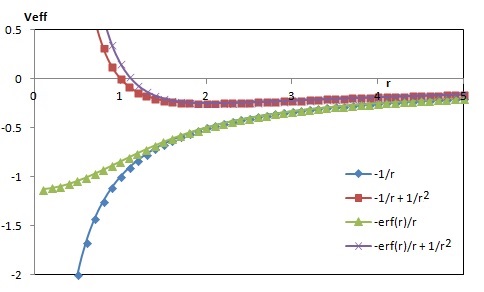}
  \caption{Comparison between the radial potentials $-1/r$ and $-\mathrm{erf}(r)/r$ in the Hydrogen atom, when $l=0$ (diamonds and squares, respectively) and $l=1$ (triangles and stars).}
  \label{fig_explain}
\end{figure}

\begin{figure}[h]
   \includegraphics[width=0.5\textwidth]{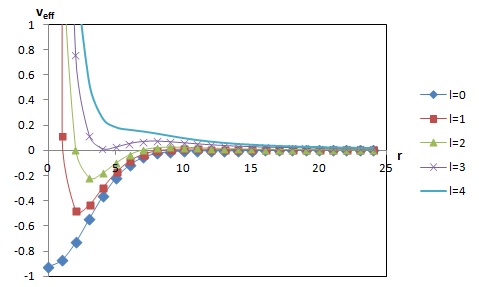}
\caption{The effect of the centrifugal term on the radial potential in the
$\mu \to 0$ limit for the fitted erfgau interaction [Eq.~(\ref{eq_fit})].}
\label{fig_centrifugal0}
\end{figure}

\begin{figure}
\includegraphics[width=0.5\textwidth]{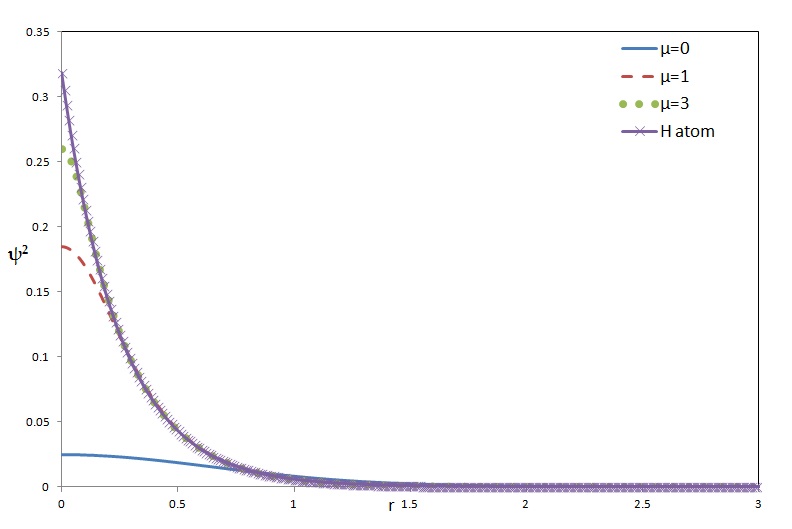}
\includegraphics[width=0.5\textwidth]{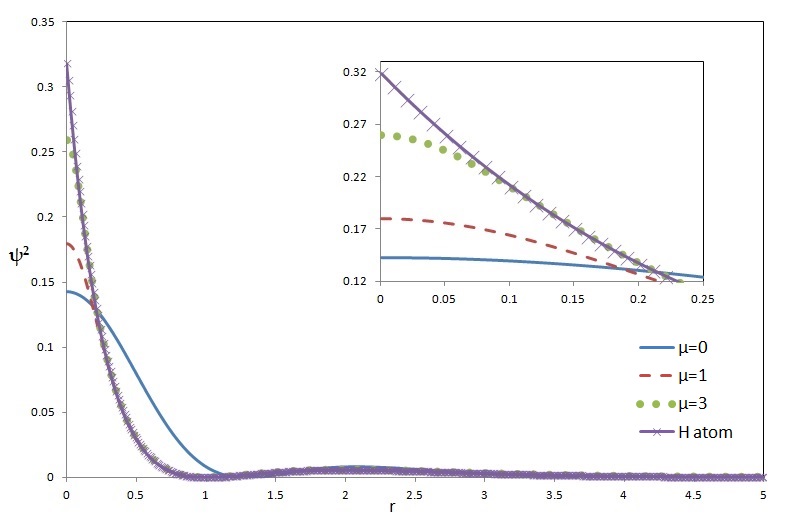}
\caption{The orbital densities $|\psi_{1s}(r)|^2$(top) and
$|\psi_{2s}(r)|^2$(bottom) derived from the model Hamiltonian of Hydrogen, 
Eq.~(\ref{hamiltonian}) with $c=0.923+1.568\,\mu$, $\alpha=0.241+1.405\,\mu$, using
different values of $\mu$.}
\label{fig_wfns}
\end{figure}

As expected from the asymptotic analysis, better results are obtained for
higher angular momentum because the centrifugal barrier, $l (l+1)/ 2r^2$,
pushes the electron away from the nucleus, into a region where the
difference between the model potential and the Coulomb potential is
negligible (see Figures~\ref{fig_explain} and ~\ref{fig_centrifugal0}).  On the other hand, when
$\mu$ is close to zero, the eigenvalues from the model potential are very
poor, because the short-range Gaussian term cannot bind an electron when the
angular momentum is too high.  When we look at the wavefuntion, for example,
the 1s and 2s orbitals (Fig.~\ref{fig_wfns}), we see that even though the
eigenvalues are very similar, the eigenfunctions can be quite different.

How important is this difference? Is the perturbation still small if
spectrum is nearly reproduced?  Moreover, can we use the same approach for other types of interactions, e.g.  a repulsive potential?  There are several ways to assess the transferability of our model potentials.  Below we use the same replacement for the electron-electron repulsion in two model systems, Harmonium and the uniform electron gas.

\section{Harmonium}

To explore whether the model potentials can be used to describe repulsive
interactions, we consider a system of two interacting electrons confined in
a harmonic oscillator potential, called Harmonium
\cite{taut,KarwowskiHarm2003}.  The Hamiltonian of
Harmonium is:
\begin{equation}
\hat{H}^\mathrm{h}(\mathbf{r}_1,\mathbf{r}_2)=-\frac{1}{2} \nabla^2_1+
\frac{\omega^2\,r^2_1}{2}-\frac{1}{2}\nabla^2_2+\frac{\omega^2\,r^2_2}{2}
+\frac{1}{|\mathbf{r}_1-\mathbf{r}_2|},
\end{equation}
where the superscript $\mathrm{h}$ stands for {\em harmonium}.
This Hamiltonian is separable if one rewrites it in terms of the center of
mass and the relative coordinates
\begin{equation}
\mathbf{R}=\frac{1}{2}(\mathbf{r}_1+\mathbf{r}_2),\quad \mathbf{r} 
=\mathbf{r}_1-\mathbf{r}_2.
\end{equation}
In the new coordinates 
\begin{equation}
\hat{H}^\mathrm{h}(\mathbf{r}_1,\mathbf{r}_2)=
\hat{H}_\mathbf{r}^\mathrm{h}(\mathbf{r})+
\hat{H}_\mathbf{R}^\mathrm{h}(\mathbf{R}), 
\end{equation}
where
\begin{align}
\hat{H}_\mathbf{r}^\mathrm{h}(\mathbf{r})&=-\nabla^2_\mathbf{r}+
\frac{\omega^2\,r^2}{4}+\frac{1}{r},\\
\hat{H}_\mathbf{R}^\mathrm{h}(\mathbf{R})&=
-\frac{1}{4}\nabla^2_{\mathbf{R}}+\omega^2\,R^2 
\end{align}
and the Schr\"odinger equation separates into two equations:
\begin{equation}
\label{harm1}
\hat{H}_\mathbf{r}^\mathrm{h}(\mathbf{r})\,
\Phi_{nlm}(\mathbf{r})=\epsilon_{nl}\,\Phi_{nlm}(\mathbf{r})
\end{equation}
and
\begin{equation}
\label{harm2}
\hat{H}_\mathbf{R}^\mathrm{h}(\mathbf{R})\,\xi_{\nu\lambda\mu}(\mathbf{R})=
\eta_{\nu\lambda}\,\xi_{\nu\lambda\mu}(\mathbf{R}),
\end{equation}
where $n,l,m$ and $\nu,\lambda,\mu$ are quantum numbers and the total energy is equal to $E_{\nu\lambda;nl}=\eta_{\nu\lambda}+\epsilon_{nl}$.  

In the case of Harmonium we can use the same approximation for the Coulomb potential as we did for the Hydrogen atom but now, instead of the {\em attractive} Coulomb interaction we have the {\em repulsive} one.
Thus, the modified Hamiltonian for the relative motion of two electrons reads 
\begin{equation}
\label{harm-erfgau}
\hat{H}_{\mathbf{r};\mu}^\mathrm{h}(\mathbf{r})=
-\nabla^2_\mathbf{r}+\frac{\omega^2\,r^2}{4}+V_\mu(r).
\end{equation}
Due to the spherical symmetry, it is convenient to express the solutions of
Eqs.~(\ref{harm1}) and (\ref{harm2}) in spherical coordinates. 
In particular, if we set 
\begin{equation}
\Phi_{nlm}(\mathbf{r}) = 
\frac{1}{r}\,\phi_{nl}(r)\,Y_{lm}(\hat{r}),
\end{equation}
where $Y_{lm}(\hat{r})$ are spherical harmonics, then in the case 
of Eq.~(\ref{harm1}) with the modified Hamiltonian (\ref{harm-erfgau}) 
we have
\begin{equation}
\label{eq_harm}
\left[-\frac{d^2}{dr^2}+\frac{l(l+1)}{r^2}+\frac{\omega^2\,r^2}{4}+
V_\mu(r)\right]\phi_{nl}(r)=\epsilon_{nl}\,\phi_{nl}(r).
\end{equation}

\subsection{How Harmonium is computed}

In order to assess the model potentials for Harmonium we solved
Eq.~(\ref{eq_harm}) numerically. To this end we discretized this
equation on a grid of $N$ equidistant points for $r\in[0,a]$ with 
the boundary conditions $\phi_{nl}(0)=\phi_{nl}(a)=0$ and the 
approximation 
\[
\frac{d^2}{dr^2}\,\phi_{nl}(r)\approx
\frac{1}{h^2}\left[\phi_{nl}(r-h)-2\phi_{nl}(r)+\
\phi_{nl}(r+h)\right], 
\]
where $h=a/N$. The discretized equation may be written as 
\[
\sum_{j=0}^N\left(A_{ij}-
\epsilon_{nl}\,\delta_{ij}\right)\phi_{nl}(r_j)=0,
\]
where $r_j=j\,h$, $j=0,1,\ldots,N$ and
\[
\footnotesize
A_{ij}=\left(\frac{2}{h^2}+\frac{l(l+1)}{r^2_j}+
\frac{\omega^2\,r^2_j}{4}+V_\mu(r_j)\right)\delta_{ij}
-\frac{\delta_{i,j+1}+\delta_{i,j-1}}{h^2}
\]
It has been solved using standard LAPACK subroutines. In the calculations we set $N=10\,000$ and $a=10/\omega$. In order to discard errors due to the numerical procedure, we computed and compared the eigenvalues of two states for which we know the analytic solution:

\begin{equation}
\small
\phi_1\,\sim\,r^{l+1}\left(1\,+\,\omega\,r\right)\mathrm{e}^{-\omega\,r^2/4}, \; \; \omega=\frac{1}{2(l+1)},\;\;\;E_1=\omega\left(l+\frac{5}{2}\right)
\end{equation}
and 
\begin{align}
\small
\phi_2\,\sim\,r^{l+1}\left[1\,+\,\frac{r\left(1+\omega\,r\right)}{2(l+1)}\right]\mathrm{e}^{-\omega\,r^2/4}, \;\; \\ \nonumber \omega=\frac{1}{2(4l+5)},\;\;E_2=\omega\left(l+\frac{7}{2}\right).
\end{align}
For $l=0$, the percentage errors in the eigenvalues obtained with the numerical integration are $5.76\times10^{-6}$\% and $6.52\times10^{-6}$\% , respectively.

\subsection{Results}

\begin{figure}
 \includegraphics[width=0.5\textwidth]{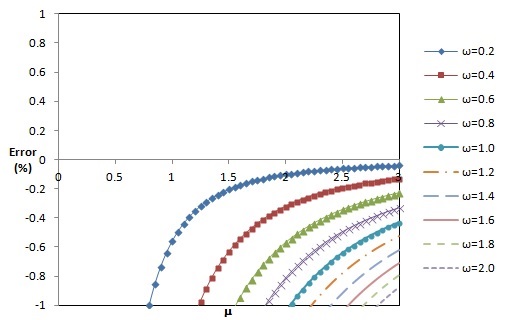}
  \includegraphics[width=0.5\textwidth]{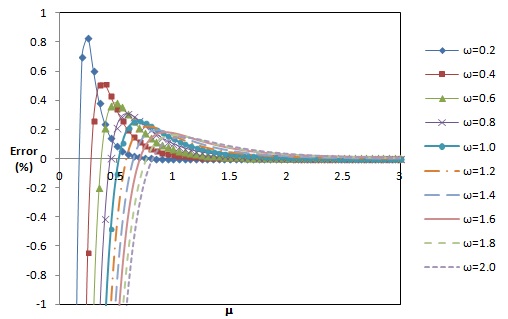}
  \includegraphics[width=0.5\textwidth]{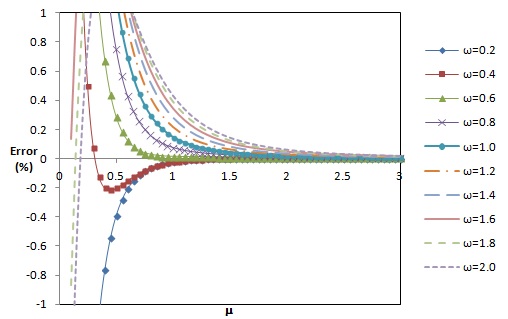}
 \caption{Errors (in \%) of the Harmonium eigenvalues as function of $\mu$, for different $\omega$. From top to bottom:a)long-range $\mathrm{erf}(\mu r)/r$, b) asymptotic (Eq.~\ref{eq_as}), and c)fitted (Eq.~\ref{eq_fit}) potentials.}
 \label{fig_harm}
\end{figure}

\begin{figure}
 \includegraphics[width=0.5\textwidth]{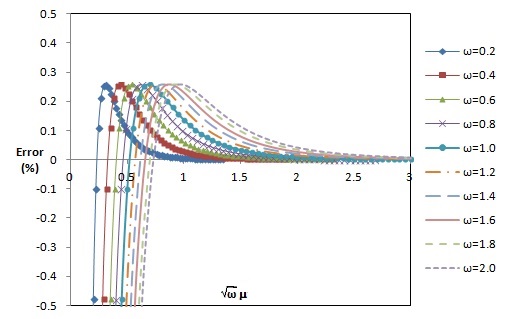}
\caption{Errors (in \%) of the Harmonium eigenvalues as function of $\sqrt{\omega} \mu$,
for different $\omega$ using the scaled asymptotic [Eq.~(\ref{eq_as})] potential.}
\label{fig_harm_scaled}
\end{figure}

\begin{figure}
 \includegraphics[width=0.5\textwidth]{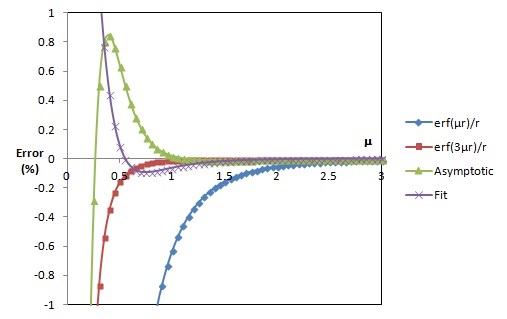}
\caption{Errors (in \%) of the excitation energy to the first excited $l=0$ state of Harmonium, as function of $\mu$, for $\omega=1$. Long-range $\mathrm{erf}(\mu
r)/r$ (diamonds), $\mathrm{erf}(3 \mu
r)/r$ (squares), asymptotic [Eq.~(\ref{eq_as})] (triangles), and fitted
[Eq.~(\ref{eq_fit})] (stars) potentials.}
\label{fig_harm-as-excit}
\end{figure}
In figure ~\ref{fig_harm}, the eigenvalues of Harmonium with $n=1$ and $l=0$
are shown for the long-range, asymptotic, and fitted potentials.  Similar to the Hydrogen atom, a deterioration at small $\mu$ is observed, but now the
asymptotic potential is slightly better than the fit to the hydrogenic
spectrum, and both are much better than the uncorrected $\mathrm{erf}(\mu
r)/r$ potential.  As the harmonic confinement weakens ($\omega \to 0$), the
average distance between electrons increases, and the models become more
accurate because of their correct $1/r$ asymptotics.  For strongly confined
electrons ($\omega >> 1$), however, the Gaussian correction factors that
were adapted to the hydrogenic spectrum do not seem appropriate for
modelling the short-range $1/r$ interaction.  In Table~\ref{tab:1} we
collected the smallest value of $\mu$ such that the error is always less than
1\%, for the different values of $\omega$.  It is clear that as the range of
the average distance between the electrons decreases ($\omega \to \infty$),
we also need to scale the range-separation parameter ($\mu \to 0$).

In order to investigate the interplay between the strength of confinement 
and the parameters of the model potential let us scale the variable in
Eq.~(\ref{harm1}) with Hamiltonian (\ref{harm-erfgau}) to reduce the 
confinement parameter to $\omega=1$. After the substitution 
$\rho=\sqrt{\omega}\,r$. We get
\begin{equation}
\small
\left[-\nabla_\mathbf{\rho}^2+\frac{\rho^2}{4}+
\frac{1}{\sqrt{\omega}}\left(\frac{\mathrm{erf}(\tilde{\mu}\rho)}{\rho}
+\tilde{c}\,\mathrm{e}^{-\tilde{\alpha}^2\rho^2}\right)\right]
\phi_{nl}(\rho)=\tilde{\epsilon}_{nl}\phi_{nl}(\rho),
\end{equation}
where
\begin{equation}
\label{scaling-omega}
\tilde{c}=c/\sqrt{\omega},\quad\tilde{\alpha}=
\alpha/\sqrt{\omega},\quad\tilde{\mu}=
\mu/\sqrt{\omega}, \quad \tilde{\epsilon}_{nl}=\epsilon_{nl}/\omega.
\end{equation}
Thus, to compensate for changing $\omega$ we have to properly scale
parameters and multiply the potential by $\sqrt{\omega}$. In 
Fig.~\ref{fig_harm_scaled} we can see that by appropriately scaling the parameters 
of the model potential we get, for all values of $\omega$, exactly the 
same energies.

\begin{table}
\begin{center}
\caption{Smallest value of the range-parameter $\mu$ needed to obtain a
percentage error less than 1\% with the model potentials for Harmonium, for
a given value of $\omega$.}
\label{tab:1}       
\begin{tabular}{lllllllllll}
\hline\noalign{\smallskip}
$\omega$ & 0.2 & 0.4 & 0.6 & 0.8 & 1.0 & 1.2 & 1.4 & 1.6 & 1.8 & 2.0\\
\noalign{\smallskip}\hline\noalign{\smallskip}
Asymptotic & 0.2 & 0.25 & 0.35 & 0.4 & 0.45 & 0.5 & 0.5 & 0.55 & 0.6 & 0.6  \\
Fitted & 0.4 & 0.25 & 0.4 & 0.5 & 0.55 & 0.6 & 0.65 & 0.65 & 0.65 & 0.65 \\
\noalign{\smallskip}\hline
\end{tabular}
\end{center}
\end{table}

To show that the same method can be used for excited states, we computed the
excitation energy from the ground state to the first excited $l=0$ state,
using the long-range, $\mathrm{erf(3 \mu r)/r}$, asymptotic, and fitted potentials (see
Figure~{\ref{fig_harm-as-excit}}).  There is some cancellation of errors (i.e., the energy spacing is better than the absolute energy), but the results are still poor for small values of $\mu$, confirming that the parameters in the model potential should be $\omega$-dependent.  We should note, however, that the erfgau potential with fixed parameters is still much better than the raw $\mathrm{erf}(\mu r)/r$ potential.

\section{Uniform electron gas}

We now examine the effect of using a modified potential on the energy of the uniform electron gas. Consider the Hartree-Fock energy of the $N$-particle spin-unpolarized uniform electron gas confined in the volume $\Omega$ with density $\rho=N/\Omega$, in the limit where $N$ and $\Omega$ go to infinity at constant $\rho$. The one-electron reduced density matrix has the well-known form\cite{JonesMarch73}
\begin{equation}
\gamma(\mathbf{r},\mathbf{r}') = 3 \rho \frac{\sin(x) - x \cos(x)}{x^3} \, ,\quad \mathrm{with} \;  x= k_F|\mathbf{r}-\mathbf{r}'|,
\end{equation}
which is not affected when we replace the Coulomb interaction, both attractive and repulsive, with the modified interaction (Eq.(\ref{eq_pot})), because it depends only on the Fermi wavenumber $k_F=(3\pi^2 \rho)^{1/3}$. Furthermore, the compensation of the electrostatic terms is maintained (\textit{i.e.} the electrostatic contribution sums up to zero, just as in the standard case). However, the exchange energy is modified to
\begin{equation}
E_x = - \frac{1}{2} \int \int d\mathbf{r} \, d\mathbf{r}' \, \gamma(\mathbf{r},\mathbf{r}')^2 V_\mu(|\mathbf{r}-\mathbf{r}'|).
\end{equation}
Using $\int\rho \, d\mathbf{r} = N$ and transforming the variables of integration, one obtains
\begin{equation}
E_x= -\frac{6}{\pi^2} N \int_0^{\infty} x^2 \left(\frac{\sin(x) - x \cos(x)}{x^3} \right)^2 V_\mu \left(\frac{x}{k_F}\right) dx.
\end{equation}
Replacing $V_\mu$ with the erfgau form of interest to us, Eq. (\ref{eq_pot}), we can then separate the integral into two terms, the Gaussian function term and the error function term
\begin{align}
\small
\int_0^{\infty} x^2 \left(\frac{\sin(x) - x \cos(x)}{x^3} \right)^2 &V_\mu \left(\frac{x}{k_F}\right) dx =\\ \nonumber
\int_0^\infty x^2 \left(\frac{\sin(x) - x \cos(x)}{x^3} \right)^2 &c \, e^{-\alpha^2 (\frac{x}{k_F})^2} dx \\ \nonumber
+ \int_0^\infty x^2 \left(\frac{\sin(x) - x \cos(x)}{x^3} \right)^2  &\frac{\mathrm{erf}(\mu \; \frac{x}{k_F})}{\frac{x}{k_F}} dx.
\end{align}

Both integrals can be easily evaluated using standard tools for numerical computations such as Mathematica\cite{Mathematica}.

\begin{figure}
\begin{center}
  \includegraphics[width=0.4\textwidth]{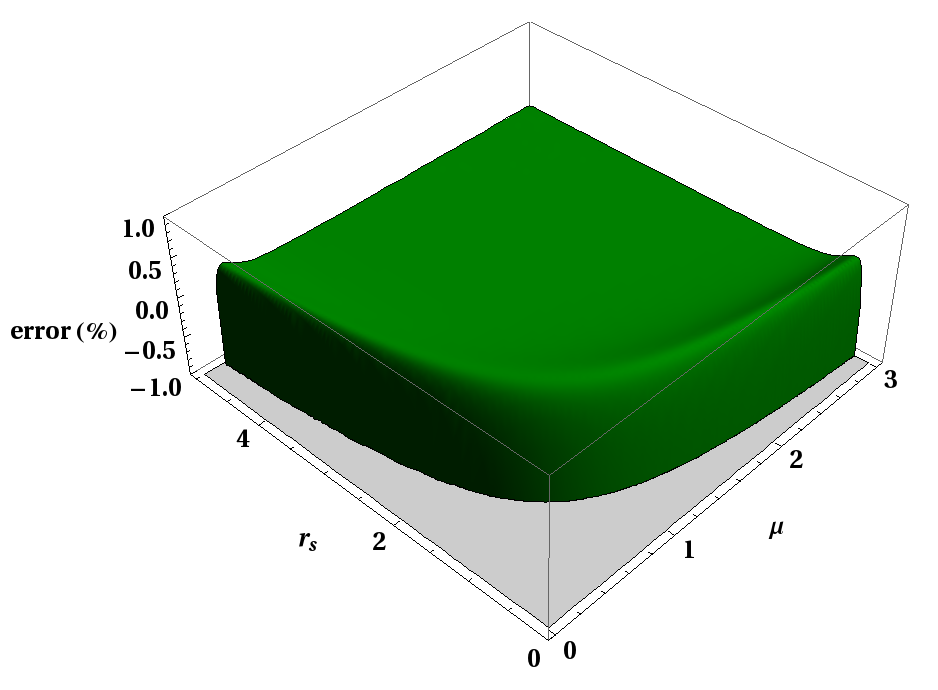}
  \includegraphics[width=0.35\textwidth]{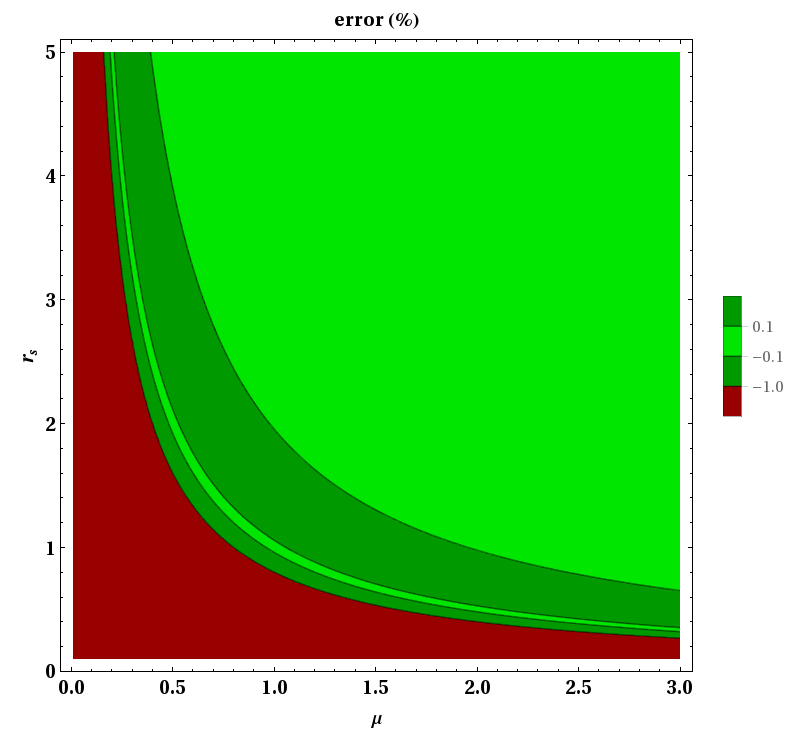}
  \caption{Error (in \%) in $\epsilon_x(r_s,\mu)$, for the uniform electron gas using the model asymptotic potential (Eq.~\ref{eq_as}).}
\label{fig_exueg}
\end{center}
\end{figure}

In Figure~\ref{fig_exueg} we show the exchange energy per particle $\epsilon_x = E_x/N$, as function of $\mu$ and the density parameter $r_s = \left(3/4 \pi \, \rho\right)^3$, using the asymptotic potential (Eq.(\ref{eq_as})). We observe that the model works well for large $\mu \,r_s$, but it does not seem possible to correct the interaction at short range.  Here, it is important to notice the similarity with Harmonium.  As $\omega$ controls the distance between the electrons, $r_s$ describes the
electron density distribution.  A small value of $\omega$ translates to
short interparticle distances, making the gas ``denser", and as
consequence, difficult to describe with the smooth potentials. This indicates that the
optimal value of $\mu$, just as in Harmonium, depends on the range of the
interaction, $r_s$, so $\mu$ should be system-dependent.

\section{Summary}

Is it possible to replace the Coulomb potential with another potential that is computationally more convenient and, if so, how should the approximate
potential be constructed?  In this work we examine if adding a Gaussian function improves the performance of the traditional $\mathrm{erf}(\mu r)/r$
potential used in range-separated DFT.  As the measure of the correctness of the model potential we have chosen the difference between the spectra of the Hydrogen atom calculated using the Coulomb potential and the modified one. 
It appears that for a reasonable range of parameters defining the new potential not only can the spectrum of the Hydrogen atom be accurately reproduced, but using the same potential to replace the \emph{repulsive}
Coulomb potential in the Harmonium atom gives a significant improvement over the uncorrected $\mathrm{erf}(\mu r)/r$ potential.

The remaining question is whether one could somehow correct the residual
error in the new potential.  One way to do this would be to, as in
range-separated DFT, use a correction functional for the neglected
short-range contributions to the exchange-correlation energy.  However, this
biases one's treatment towards the ground-state energy and electron density:
a different (and certainly much harder to construct) functional would be
needed to correct other properties (e.g., excited-state properties) of the
system.  There is another way, however: the results of a few calculations at
sufficiently large values of $\mu$ can be extrapolated to the physical $\mu
\to \infty$ limit.  This approach is applicable to any property, not just
those that are readily accessible from KS DFT. Furthermore, replacing the
electron-electron repulsion potential with a smooth function has major
computational advantages, as it allows one to use smaller basis sets, with
fewer polarization functions.

There are also cases where it may be favorable to replace the  Coulombic
electron-nuclear interaction with a model potential like those considered in
this paper.  For example, these smoothed Coulomb potentials could be used,
instead of pseudopotentials, for diffusion quantum Monte Carlo and
plane-wave DFT calculations.  In those cases, the procedure would be the
same: the system would be solved for several choices of the smoothed
electron-nuclear interaction, and the results then extrapolated to the
physical $\mu \to \infty$ limit.

\begin{acknowledgements}
CEGE acknowledges CONACyT and the Secretariat of Innovation, Science and
Technology of the State of Morelos for the scholarship for graduate studies,
and MITACS for funding her visit to the Laboratoire de Chimie Th\'eorique,
where this work was initiated.  PWA acknowledges support from NSERC and
Compute Canada.

This paper is dedicated to Alberto Vela, on the occasion of his sixtieth birthday. Alberto likes to improve results through understanding, rather than haphazardly. We find this trait admirable and worthy of imitation, and hope that through this paper we provide a better understanding of how to model electron interactions.

\end{acknowledgements}

\begin{appendix}

\section{Hydrogenic atoms: $\mu$ dependence and integrals' scaling.}

The leading term in the hydrogenic radial function is
\begin{equation}
R_{nl}(r)^2 \sim Z^3 (Zr)^{2l} \mathrm{e}^{-2Zr/n},
\label{app_eq_leading}
\end{equation}
therefore 
\begin{equation}
R_{nl}(r)^2 r^2 dr \sim \rho^{2l+2} \mathrm{e}^{-2\rho/n}d\rho
\end{equation}
where $\rho = Zr$. Now, let
\begin{equation}
I(\mu)=\int_0^\infty f(\mu r) R_{nl}(r)^2 r^2 dr
\end{equation}
then
\begin{align}
I(\mu) &\sim \int_0^\infty f \left(\frac{\mu \rho}{Z} \right) \rho^{2l+2} \mathrm{e}^{-2\rho/n} d\rho \nonumber \\
 &= \left(\frac{Z}{\mu} \right)^{2l+3} \int_0^\infty f(x) x^{2l+2} \mathrm{e}^{-2Zx/\mu n} dx \nonumber \\
 &= \left( \frac{Z}{\mu}\right)^{2l+3} \sum_{i=0}^{\infty} a_i \left(\frac{Z}{\mu} \right)^i ,
\end{align}
where $x=\mu \rho/Z$ and
\begin{equation}
a_i = \frac{(-1)^i}{i!} \left(\frac{2}{n}\right)^i \int_0^\infty f(x) x^{2l+i+2} dx.
\end{equation}
For $f(\mu\,r /Z) = \mathrm{erf}(\mu\,r)/Z r$, $f(x) = (\mu/Z)\mathrm{erf}(x)/x$ and the power of the asymptotic term is $(2l+2)$.

\section{Model Hamiltonian from first-order perturbation theory.}

We define the model Hamiltonian as
\begin{equation}
\hat{H}_\mu=-\frac{1}{2}\nabla^2-\left[c\,\exp(-\alpha^2r^2) + \frac{\mathrm{erf}(\mu\,r)}{r} \right],
\end{equation}
where the parameters $c$ and $\alpha$ should be chosen so that the eigenvalues 
of this operator are as close as possible to the ones of the physical operator,
\begin{equation}
\hat{H_0}=-\frac{1}{2} \nabla^2 - \frac{1}{r}.
\end{equation}
The difference between the two operators 
\begin{equation}
w_\mu=\hat{H}_\mu-\hat{H_0} = \frac{\mathrm{erfc(\mu\,r)}}{r} - 
c \exp(- \alpha^2 r^2)
\end{equation}
is treated as a perturbation.
We want the perturbation to vanish as $\mu \to \infty$. Moreover, we would
like to keep a single parameter, $\mu$, and make $c$ and $\alpha$ functions
of $\mu$.  As we mentioned in section 3.1, one way to produce 
$\langle\psi_i|w_\mu|\psi_i\rangle = 0$ is to choose $\alpha$ increasing with
$\mu$.  When we expand the expectation value of $w_\mu$ for large values 
of $\alpha$ and $\mu$ we obtain, for $l=0$ states, the integrals of the hydrogenic functions  $\psi_{n0} = \{4 e^{-r}, \; (2-r)e^{-r/2},\; \frac{4}{729} (27-18r+2r^2)e^{-r/3}\}$ are
\begin{equation}
\langle\psi_{n0}|w_\mu|\psi_{n0}\rangle=n^{-3}\left[
A(\alpha,c)+B(\mu)\right]+O(\mu^{-5}),
\label{expansion}
\end{equation}
where
\[
A(\alpha,c)=-\frac{c\,\sqrt{\pi}}{\alpha^3}+\frac{4c}{\alpha^4}
\]
comes from $\langle\psi_{n0}|c\,\exp(-\alpha^2 r^2)|\psi_{n0}\rangle$ and
\[
B(\mu)=\frac{1}{\mu^2}-\frac{8}{3\sqrt{\pi}\mu^3}+\frac{3}{2\mu^4},
\]
from $\langle \psi_{n0}|\mathrm{erfc}(\mu r)/r|\psi_{n0}\rangle$. Note that we used arbitrary multiplicative factors (4, 1 and $\frac{4}{729}$ respectively) to simplify the expressions. We can use this trick because we want to equate all expressions to 0.

In order to eliminate terms of order $\mu^{-2}$, we set  
$\sqrt{\pi}\,c\,\alpha^{-3}=\mu^{-2}$. This means that $c=\gamma\,\mu$,  
$\alpha=\kappa\,\mu$ and  $\gamma=\kappa^3/\sqrt{\pi}$. 
Substituting these values into the expansion (\ref{expansion}) corresponding
to $n=1$ we see that the coefficient of $\mu^{-3}$ vanishes if
$\gamma=(2\,\kappa^4)/(3\sqrt{\pi})$. From the last two equations we get 
\[
\gamma=\frac{27}{8 \sqrt{\pi}},\;\;\;\;\kappa=\frac{3}{2} 
\]
for which the energy of the $S$ states is correct up to $\mu^{-4}$.  To
eliminate the error for $\mu^{-4}$, we would need to consider corrections from second-order
perturbation theory. For $l=1$ the expansion of the expectation value
of $w_\mu$ starts with terms proportional to $\mu^{-4}$ and, in general, for
an arbitrary $l$, the leading term of the expansion is proportional to
$\mu^{-(2l+2)}$.

Now let us consider the expansion of the expectation values of $w_\mu$ for
the linear forms $\alpha = \kappa \mu + \alpha_0$ and $c = \gamma \mu + c_0$ with
the parameter $\gamma$ and $\kappa$ from the previous step.  The condition
for the elimination of the $\mu^{-2}$ term remains the same as above. The coefficient of $\mu^{-3}$ is equal to
\begin{equation}
\label{m3}
d_3=-\frac{8\,c_0\,\sqrt{\pi}}{27}+2\,\alpha_0
\end{equation}
and is the same for all $l=0$ states. The coefficient of $\mu^{-4}$ for $l=0$ states is $n$-dependent. For $n=1$ it is equal to
\begin{equation}
\label{m4}
d_4=\frac{1}{6}+\frac{64\,c_0}{81}-\frac{64\,\alpha_0}{9\,\sqrt{\pi}}+
\frac{16}{27}\,\sqrt{\pi}\,c_0\,\alpha_0-\frac{8\,\alpha_0^2}{3}
\end{equation} 
Solving equations $d_3=d_4=0$ for $c_0$ and $\alpha_0$ we obtain: 
\begin{equation}
c_0 =\frac{27\,\alpha_0}{4\,\sqrt{\pi}}, \quad \quad 
\alpha_0 = \frac{8 \pm \sqrt{64 - 18 \pi}}{12 \sqrt{\pi}}.
\label{eq_pt-lineal}
\end{equation}
This gives two possible solutions: $\{ c_0= 0.94364, \alpha_0=0.24778 \} $
and $\{ c_0= 1.92115, \alpha_0=0.50446 \}$.  When we use either of these sets
of parameters and expand the first-order correction to fourth order in
$1/\mu$, we have:
\[
1s:0,\;\;\;\;2s:-\frac{1}{48 \mu^4},\;\;\;\;3s:-\frac{2}{81\mu^4},\;\;\;\;
2p:\frac{1}{6\mu^4},\;\;\; \]
\[
3p:\frac{1}{9\mu^4},\;\;\;\;3d:0.
\] 
Thus the error of the first-order correction to the eigenvalues 
is proportional to $\mu^{-4}$.

\end{appendix}

\noindent
\bibliographystyle{spmpsci}      

\begin{thebibliography}{}


\bibitem{KS65}
W. Kohn, and L. J. Sham, Self-consistent equations including exchange and correlation effects, Phys. Rev., 140, A1133 (1965)

\bibitem{ParrYang}
R. G. Parr, W. Yang, Density-Functional Theory of Atoms and Molecules. Oxford: Oxford University Press (1994).

\bibitem{LDA80}
D. M. Ceperley and B. J. Alder, Ground State of the Electron Gas by a Stochastic Method, Phys. Rev. Lett. 45, 566 (1980) 

\bibitem{LDAPZ}
J. P. Perdew and A. Zunger, Self-interaction correction to density-functional approximations for many-electron systems, Phys. Rev. B 23, 5048 (1981).

\bibitem{LDAPW}
J. P. Perdew and Y. Wang, Accurate and simple analytic representation of the electron-gas correlation energy, Phys. Rev. B 45, 13244 (1992).

\bibitem{HaJo74}
J. Harris, and R. O. Jones, The surface energy of a bounded electron gas, J. Phys. F: Metal Phys., 4, 1170-1186 (1974).

\bibitem{LanPer75}
D. C. Langreth, and J. P. Perdew, Solid State Commun., The exchange-correlation energy of a metallic surface, Solid Sate Commun., 17, 1425-1429 (1975).

\bibitem{GunnLund76}
O. Gunnarsson, and B. I. Lundqvist, Exchange and correlation in atoms, molecules and solids by Spin-Density-Functional formalism, Phys., Rev. B, 13, 4274-4298 (1976). 

\bibitem{ACYang98}
W. Yang, Generalized adiabatic connection in density functional theory, J. Chem. Phys., 109, 10107-10110 (1998).
\bibitem{Seidl2003}
M. Seidl, Density functional theory from the extreme limits of correlation, Int. J. Quantum Chem, 91, 145 (2003).

\bibitem{MoriCohenYang06}
P. Mori-S\'anchez, A. J. Cohen and W. Yang, Self-interaction-free exchange-correlation functional for thermochemistry and kinetics, J. Chem. Phys., 124, 091102 (2006).

\bibitem{CohenMoriYang07}
A. J. Cohen, P. Mori-S\'anchez, and W. Yang, Assessment and formal properties of exchange-correlation functionals constructed from the adiabatic connection, J. Chem. Phys., 127, 034101 (2007).

\bibitem{ColonnaSavin99}
F. Colonna and A. Savin, Correlation energies for some two- and four- electron systems along the adiabatic connection in density functional theory, J. Chem., Phys., 110, 2828 (1999).

\bibitem{PoColLeinStollSavin03}
R. Pollet , F. Colonna, T. Leininger, H. Stoll, H.-J. Werner, and A. Savin, Exchange-correlation energies and correlation holes for some two- and four-electron atoms along a nonlinear adiabatic connection in density functional theory, Int. J. Quantum Chem., 91 (2), 84-93 (2003).

\bibitem{TeaCorHel09}
A. M. Teale, S. Coriani, and J. Helgaker, The calculation of adiabatic-connection curves from full configuration-interaction densities:Two-electron systems, J. Chem. Phys., 130, 104111 (2009).

\bibitem{TeaCorHel10}
A. M. Teale, S. Coriani, and J. Helgaker, Range-dependent adiabatic connections, J. Chem. Phys., 133, 164112 (2010). 

\bibitem{ValBer89}
A. Valance, H Bergeron, Model potentials or pseudopotentials: the connection via supersymmetry, J. Phys. B, Mol. Opt., Phys., 22, L65-L69 (1989).

\bibitem{Lep97}
G. P. Lepage, How to renormalize the Scr\"odinger equation, J.C.A. Barata, A.P.C. Malbouisson, S.F. Novaes (Eds.), Particles and Fields,
Proceedings of the Ninth J.A. Swieca Summer School, World Scientific, Singapore, (1998).

\bibitem{SavinColPol03}
A. Savin, F. Colonna and R. Pollet, Adiabatic connection approach to density functional theory of electronic systems, Int. J. Quantum Chem., 93, 166-190 (2003).

\bibitem{TouSavinFlad}
J. Toulouse, A. Savin, and H.-J. Flad, 
Short-Range exchange-correlation energy of a uniform electron gas with 
modified electron-electron interaction, Int. J. Quantum Chem. 
100, 1047-1056 (2004).

\bibitem{taut}
M. Taut, Two electrons in an external oscillator potential: Particular
analytic solutions of a Coulomb correlation problem, Phys. Rev. A
\textbf{48}, 3561-3566 (1993).

\bibitem{KarwowskiHarm2003}
J. Karwowski, and L. Cyrnek, Two interacting particles in a parabolic well:harmonium and related systems, Computational Methods in Science and Technology, 9(1-2), 67-78 (2003).


\bibitem{JonesMarch73}
W. Jones and N. H. March, Theoretical Solid State Physics, vol. 2. Dover, New York (1973).

\bibitem{Mathematica}
Wolfram Research, Inc., Mathematica, Version 11.0, Champaign, IL (2016).

\end{thebibliography}


\end{document}